\documentclass{article}
\usepackage{spconf,amsmath,graphicx}

\usepackage{booktabs}
\usepackage{multicol, multirow}
\usepackage{tabularx, makecell, enumitem}
\usepackage{amssymb}
\usepackage{adjustbox}
\usepackage[dvipsnames]{xcolor}
\usepackage{pifont}

\usepackage[ruled,vlined]{algorithm2e}

\title{Are soft prompts good zero-shot learners for speech recognition?}

\name{\begin{tabular}{c} 
Dianwen Ng$^{1,2}$, Chong Zhang$^{1}$, Ruixi Zhang, Yukun Ma$^{1}$, Fabian Ritter-Gutierrez$^{2}$\\ Trung Hieu Nguyen$^{1}$, Chongjia Ni$^{1}$, Shengkui Zhao$^{1}$, Eng Siong Chng$^{2}$, 
Bin Ma$^{1}$ \thanks{This work was supported by Alibaba Group through Alibaba Innovative Research (AIR) Program and Alibaba-NTU Singapore Joint Research Institute (JRI), Nanyang Technological University, Singapore}
\end{tabular}}

\address{$^1$Speech Lab of DAMO Academy, Alibaba Group \\
 $^2$School of Computer Science and Engineering, Nanyang Technological University, Singapore}
 
\begin{document}
%
\maketitle
\begin{abstract}

\end{abstract}
Large self-supervised pre-trained speech models require computationally expensive fine-tuning for downstream tasks. Soft prompt tuning offers a simple parameter-efficient alternative by utilizing minimal soft prompt guidance, enhancing portability while also maintaining competitive performance. However, not many people understand how and why this is so. In this study, we aim to deepen our understanding of this emerging method by investigating the role of soft prompts in automatic speech recognition (ASR). Our findings highlight their role as zero-shot learners in improving ASR performance but also make them vulnerable to malicious modifications. Soft prompts aid generalization but are not obligatory for inference. We also identify two primary roles of soft prompts: content refinement and noise information enhancement, which enhances robustness against background noise. Additionally, we propose an effective modification on noise prompts to show that they are capable of zero-shot learning on adapting to out-of-distribution noise environments.

\vspace{-0.3cm}
\section{Introduction}
\label{sec:intro}
While large pre-trained speech models \cite{baevski2020wav2vec, ng2020disentangling, radford2023robust} like HuBERT \cite{hsu2021hubert} and WavLM \cite{chen2022wavlm} have adopted an inherent understanding of speech and context, it is nevertheless essential to fine-tune these models to the downstream tasks. Full fine-tuning of the model is computationally expensive and does not scale well, requiring training millions of parameters for each downstream application. Additionally, the heavy cost of storing these large models for every individual task compromises their deployment and accessibility.

Soft prompt tuning \cite{lester2021power, li2021prefix, kim2023prompt} is a parameter-efficient tuning (PET) method that mitigates the limitations by calibrating a small number of trainable parameters corresponding to a fixed number of latent tokens referred to as ``\textit{prompts}". The soft prompts are prepended to the input embeddings to provide guiding signals for the model to better understand the task. The pre-trained model is kept completely frozen during this process, which makes it extremely portable than full fine-tuned parameter weights. Besides, \cite{chen2023exploring, chang2023speechprompt, guo2023prompttts} have demonstrated that such a method is competitive or even performs better than the full fine-tuning model.
Although prompt tuning has become a popular method for fine-tuning large pre-trained models, the precise mechanisms through which they enhance model performance remain an open question. 

In \cite{oymak2023role}, it suggests that soft prompts assist in accelerating convergence and enhance attention. \cite{lester2021power} argues that they may resemble natural language tokens, similar to hand-engineered prompts. However, \cite{bailey2023soft} found that soft prompts do not generally correspond to natural language prompts. Notably, many of these explorations were conducted in the context of NLP and might not directly apply to speech models, rendering them less accountable. This motivates us to investigate the role of soft prompts within the context of the automatic speech recognition (ASR) task. Specifically, \textbf{\textit{what do the soft-prompts mean to the model and how can we effectively exploit them?}} Our key findings are summarized as follows: 
\begin{itemize}[leftmargin=*] \vspace{-0.1cm}
    \item We show that soft prompts can yield better performance, but the model is susceptible to bad modifications on the prompts, which could lead to increase recognition errors. However, we find that they are not necessarily required for inference, as the model can still perform well without them. 
    \vspace{-0.5cm}
    \item We identified two primary roles of soft prompt: content refinement and augmenting noise information in the latent speech representations to improve robustness against background noise distortion in speech signals. \vspace{-0.1cm}
    \item We demonstrate an effective modification to the noise prompts to achieve zero-shot learning on adapting to out-of-domain noise environments.
\end{itemize}


\section{Empirical characterization of the role of soft-prompts}
\label{sec:method}
In this section, we illustrate the network computation involved in soft-prompt tuning and outline the goals of this paper, which include the experimental setups.

\subsection{Network Computation of Prompt Tuning}
Let $\mathbf{P} \in \mathbb{R}^{m \times d}$ represent the trainable prompts comprising $m$ tokens, each with a dimension of size $d$, following the acoustic features denoted by $\mathbf{X} \in \mathbb{R}^{T \times d}$. To incorporate these prompts into our networks, we prepend them to $\mathbf{X}$, resulting in the augmented matrix $\mathbf{X_P} := [\mathbf{X}, \mathbf{P}] \in \mathbb{R}^{(T+m) \times d}$. This new matrix, $\mathbf{X_P}$, serves as the latent input for the transformer blocks. Then, consider a single-head attention layer, the output of the attention with prompt tuning is presented by \begin{equation}
    \mathcal{O} = \varphi\Big(\frac{1}{\sqrt{d}}\mathbf{X_PW_\mathrm{Q}W_\mathrm{K}^\intercal X_P^\intercal} \Big)\mathbf{X_PW}_V
\end{equation} where $\mathbf{W}_Q$, $\mathbf{W}_K$ and $\mathbf{W}_V$ are the frozen pre-trained weights for \textit{query}, \textit{key} and \textit{value}. $\varphi$ denotes the softmax nonlinearity function that acts row-wise for a $(T+m) \times (T+m)$ matrix. We generalize this to the multi-headed attention layer and the output of the attention layer is scalarized with $\mathbf{W}_O$ as in \begin{equation}
    \text{MultiHead} = \text{Concat}\big( \mathcal{O}_1, \mathcal{O}_2, \ldots, \mathcal{O}_h\big)\mathbf{W}_O
\end{equation} where $\mathcal{O}$ is split into $h$ heads, each with a dimension of $d/h$. 

Here, \textbf{\textit{soft-prompt tuning}} exclusively occurs within the self-attention module, providing an additive effect on the output representations. The latent features would attend to the prompt vectors through the behavior of the pre-trained self-attention module to receive an extra guiding signal that refines the representations for the downstream task. 

Our goal is to investigate the impact of this tuning approach and empirically characterize the role of these prompt vectors in the context of automatic speech recognition (ASR) with training and testing corpus that may contain some background noise, as often encountered in real-world scenarios. We begin by evaluating the effectiveness of prompt tuning compared to the baseline, where the entire network is frozen and uses the same pre-trained speech encoder. We then seek to identify the factors contributing to the performance difference. Following that, we dissect the individual prompt vectors to analyze their role in fine-tuning the latent representations, revealing surprising and intriguing findings that allow for a simple yet effective modification of prompts that demonstrate the zero-shot learning potential in ASR domain adaptation.

\subsection{Training Details}
We employ the HuBERT \cite{hsu2021hubert} encoder, which trains on clean audio speech, as the backbone of our self-supervised pre-training model. Rather than choosing other noise-robust candidates, we particularly select this model to examine how the soft prompts adapt their trainable prompt vectors to provide instructions for handling noisy speech. This approach allows us to explore the prompts' ability to manage the adverse influence of noise independent of the noise-robust capability of the pre-training backbone. In our work, we mainly scrutinize the behavior of prompt tuning with $m=20$ tokens, as they are relatively easier to manage. However, we also assess the generalization performance with a bigger prompt's size of $m=50$ and determine the potential gain in accuracy with increased prompt complexity. All models, including the baseline, are fine-tuned with the SUPERB \cite{yang2021superb} benchmark decoder. The learning rate is chosen based on grid-searched within the range [2e-5, 3e-4], with 1e-4 being the best on LibriSpeech's dev-clean set. The remaining fine-tuning configurations follow the 100h configs provided by FairSeq. Note that all table results are obtained without using LM.

\subsection{Datasets}
We trained our models using 100h of synthesized noisy LibriSpeech \cite{panayotov2015librispeech} data, where we corrupt the utterance with a sampled noise ranging from 0 to 20 dB at 80\% of the time. The noise dataset comes from FreeSound \cite{font2013freesound}, comprising 16 kHz of noise data categorized into stationary (Type A) and non-stationary (Type B) noise. Type A includes Car, Metro, and Traffic noises, whereas Type B consists of Babble, Airport/Station, Cafe, and AC/Vacuum noises. Each type of noise had 10 and 8 different audio streams in the training and evaluation sets, respectively, resulting in around 2 hours of noise audio. We evaluate the performance of all models on the official test-clean and test-other LibriSpeech datasets without any additional noise augmentation. In addition, we assess the performance on noisy ASR (test-noisy), which includes pre-mixed noisy LibriSpeech data ranging from 0 to 20 dB. This test-noisy dataset comprised a total of 4,200 instances of noisy test data. The noise data and pre-mixed noisy test sets are open sourced \cite{prasad2021investigation}. \vspace{-0.2cm}

\section{Analytical Results} \vspace{-0.1cm}
\label{sec:result}
\subsection{Performance: HuBERT \ding{100} vs. Prompt Tuning}
Before analyzing the prompt vectors to understand the relationship of the soft-prompts to refining the representations for ASR, we measure the performance gain achieved by introducing soft-prompt tuning vs. frozen (\ding{100}) vanilla HuBERT. Additionally, we assess the model's sensitivity by replacing the learned soft-prompts with random standard Gaussians.

\begin{table}[!ht]
\centering\footnotesize
\tabcolsep=0.15cm
\renewcommand{\arraystretch}{1.35}
\vspace{-0.3cm}
\caption{Performance comparison between the listed models on the LibriSpeech test set, where the noisy set contains a subset of clean speech corrupted with SNR noise of 0-20dB.}
\begin{tabular}{l|c|ccc}
\Xhline{2\arrayrulewidth}
\multirow{2}{*}{Methods} & \multirow{2}{*}{\begin{tabular}[c]{@{}c@{}}No. of\\ Prompts\end{tabular}} & \multicolumn{3}{c}{WER (\%) of LibriSpeech Test} \\ \cline{3-5} 
                       &                     & \multicolumn{1}{c|}{Clean} & \multicolumn{1}{c|}{Other} & Noisy \\ \hline
HuBERT \ding{100} (Base)        & N.A.                & \multicolumn{1}{c|}{7.09}  & \multicolumn{1}{c|}{17.61} & 29.04 \\ \hline
Prompt Tuning          & \multirow{2}{*}{20} & \multicolumn{1}{c|}{6.64}  & \multicolumn{1}{c|}{16.30} & 27.06 \\
\ding{229} Replace Random Prompts &                     & \multicolumn{1}{c|}{8.27}  & \multicolumn{1}{c|}{22.63} & 43.69 \\ \hline
Prompt Tuning          & \multirow{2}{*}{50} & \multicolumn{1}{c|}{6.37}  & \multicolumn{1}{c|}{15.99} & 26.13 \\
\ding{229} Replace Random Prompts &                     & \multicolumn{1}{c|}{9.83}  & \multicolumn{1}{c|}{28.81} & 54.80 \\ \Xhline{2\arrayrulewidth}
\end{tabular}
\label{tbl:table1}
\end{table}
\vspace{-0.2cm}
In Table \ref{tbl:table1}, we observe that prompt tuning demonstrates better generalization to the ASR task than frozen HuBERT, resulting in reduced WER scores across all clean, other, and noisy test set. The performance improvement becomes more pronounced as we increase the prompt token size, suggesting that a larger number of prompt tokens might provide more information for fine-tuning contextual representations. Furthermore, we note that the model is strongly influenced by the prompts. This is evident from the substantial performance degradation observed when random prompts are swapped in. This observation underscores the model's susceptibility to adversarial attacks and highlights the importance of prompt integrity and design in maintaining robustness.

\begin{figure}
    \centering
    \includegraphics[width=0.75\linewidth]{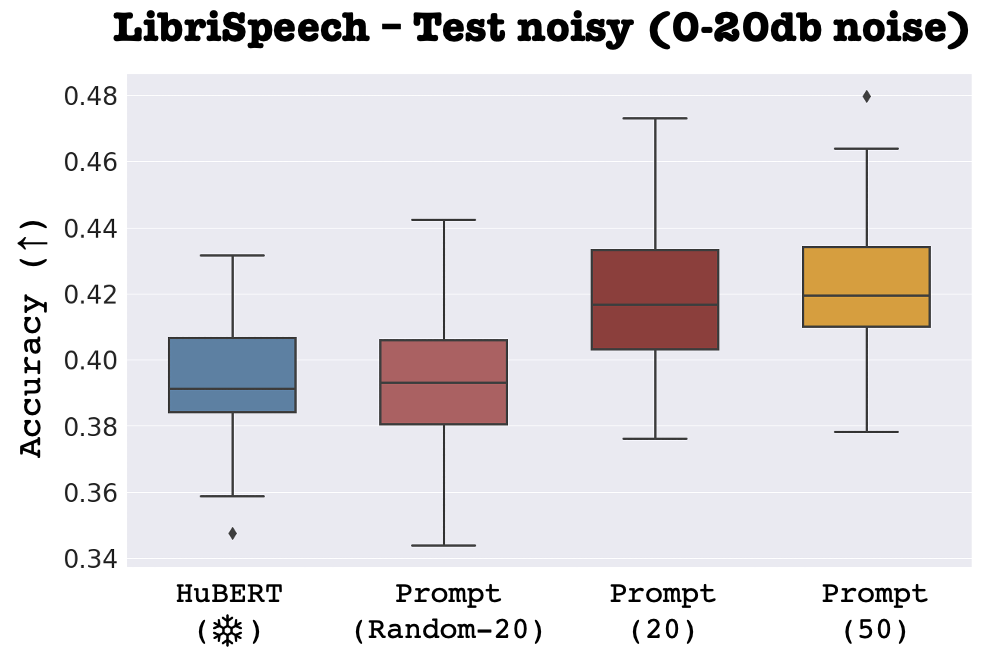}
    \vspace{-0.2cm}
    \caption{Boxplot of accuracy in classifying noisy speech into their mixed noise types using the model's pooled latent representations that carry embedded noise information.}
    \label{fig:base1}
    \vspace{-0.4cm}
\end{figure}

Next, following the discussion in \cite{gong2023whisper}, we asked a similar question: \textbf{Is the improved performance of prompt tuning in noise robustness the result of making the representations \textit{noise-variant} or \textit{noise-invariant?}} Given that we have access to the labels of the corrupted noise types in Test-noisy, we performed global average pooling on the transformer (taking into account the influence of the prompts) output and conducted noise classification using a random forest. The results, depicted in Fig. 2, illustrate the accuracy obtained from the random forest. Notably, we observed that prompt tuning appears to inject noise information, leading to significantly improved accuracy in noise classification. 
Thus, prompt tuning helps to refine the representations to be more noise-variant.

\subsection{Exemplify the Roles of the Prompts}
\begin{figure}
    \centering
    \includegraphics[width=0.9\linewidth]{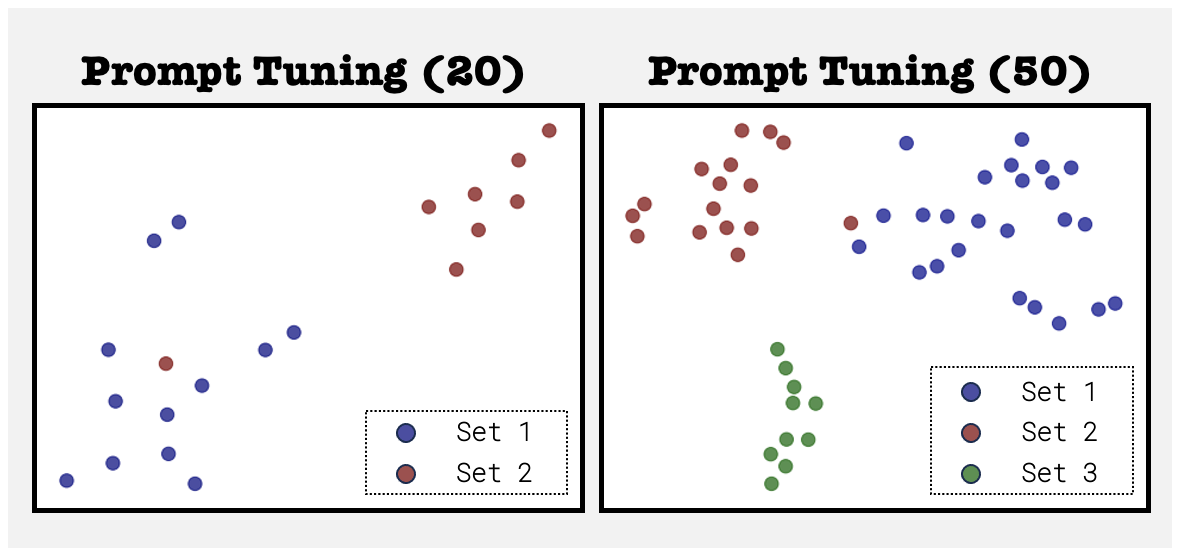}
    \vspace{-0.2cm}
    \caption{t-SNE plot of the prompt vectors, displaying the cluster relationship of each prompt in the downsampled 2D space.}
    \label{fig:role1}
    \vspace{-0.2cm}
\end{figure}

We attempted to visualize the relationship of the prompt vectors on a t-SNE plot. We found two distinctive clusters for a prompt size of 20 and three for a prompt size of 50. The color clustering in Fig. \ref{fig:role1} is obtained from K-Means of the original soft-prompts. The plot strongly suggests that the learned prompts likely correspond to 2--3 primary roles in fine-tuning the contextual representations.

\begin{figure}
    \centering
    \includegraphics[width=0.68\linewidth]{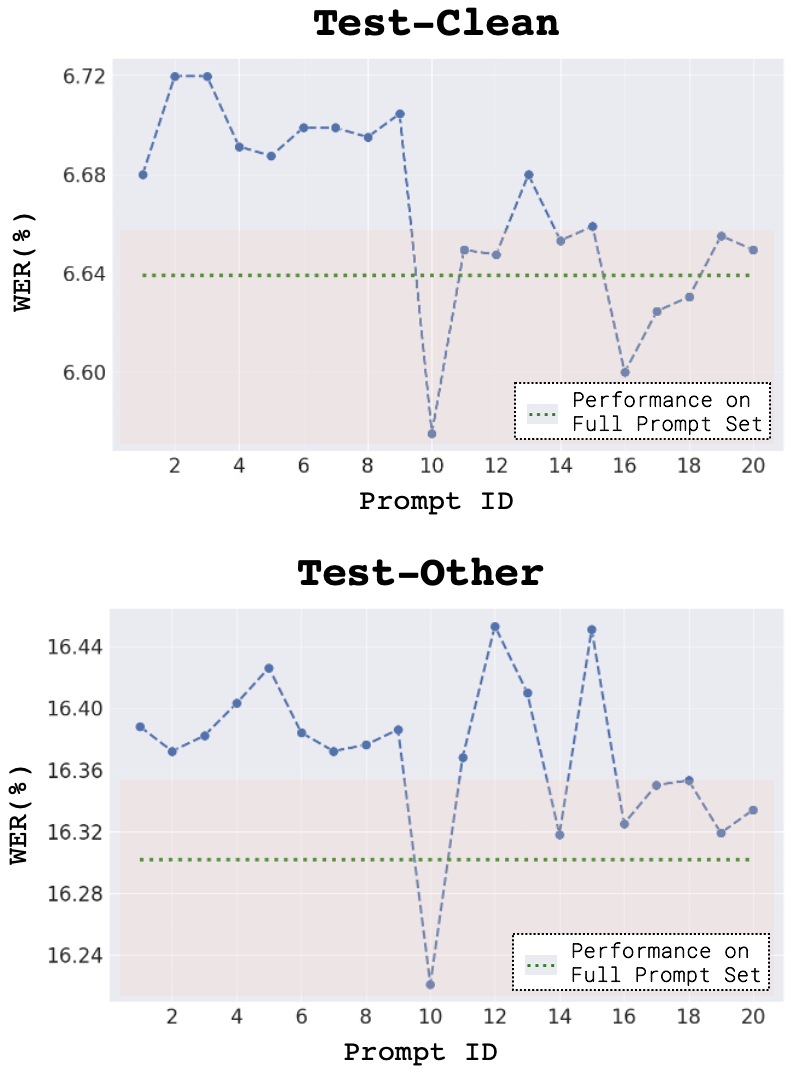}
    \vspace{-0.2cm}
    \caption{WER of prompt tuning (20) on different test set after removing a single prompt vector given from their prompt ID.}
    \label{fig:role2}
    \vspace{-0.3cm}
\end{figure}

Subsequently, we identified that the two primary roles of prompt tuning with 20 tokens are likely responsible for content tuning and noise information tuning. Specifically, we assessed the impact of deactivating each prompt on recognition performance, as depicted in Fig \ref{fig:role2}. Particularly, we observed that disabling certain prompts led to an increase in recognition errors on clean speech (i.e., containing content-rich corpus), associating their role in content refinement. Conversely, the other set of prompts within the orange zone exhibited a minor impact, suggesting a trivial role in content tuning. We later argue that this set is responsible for the noise. Note that the two prompt sets coincide with the K-Means clustering in Fig. \ref{fig:role1}. The identified prompt sets are as listed \footnote{Set 1: \{1, 2, 3, 4, 5, 6, 7, 8, 9, 12, 13, 15\}; Set 2: \{10, 11, 14, 16, 17, 18, 19, 20\}, where the indices refer to the prompt ID in Fig. \ref{fig:role2}.}.

\begin{figure}
    \centering
    \includegraphics[width=0.78\linewidth]{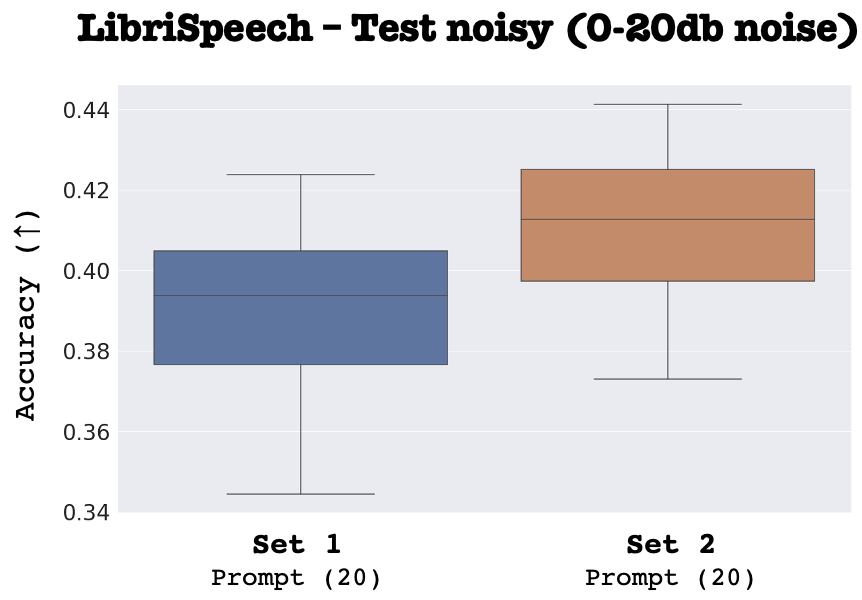}
    \vspace{-0.2cm}
    \caption{Boxplot of accuracy in classifying noisy speech into their mixed noise types using the model's pooled latent representations with different soft prompt (20) subsets $^1$.}
    \label{fig:role3}
    \vspace{-0.4cm}
\end{figure}

\begin{table}[!ht]
\centering\footnotesize
\tabcolsep=0.33cm
\vspace{-0.3cm}
\renewcommand{\arraystretch}{1.35}
\caption{Performance comparison between the prompt tuning (20) model with the utilization of different prompt sets on the LibriSpeech test set.}
\begin{tabular}{l|c|ccc}
\Xhline{2\arrayrulewidth}
\multirow{2}{*}{Methods}                                                 & \multirow{2}{*}{\begin{tabular}[c]{@{}c@{}}Prompts\\ (Set)\end{tabular}} & \multicolumn{3}{c}{WER (\%) of LibriSpeech Test}           \\ \cline{3-5} 
       &       & \multicolumn{1}{c|}{Clean} & \multicolumn{1}{c|}{Other} & Noisy \\ \hline
\multirow{3}{*}{\begin{tabular}[c]{@{}l@{}}Prompt Tuning (20)\end{tabular}} & Full                     & \multicolumn{1}{c|}{6.64} & \multicolumn{1}{c|}{16.30} & 27.06 \\
       & Set 1 & \multicolumn{1}{c|}{6.65}  & \multicolumn{1}{c|}{16.34} & 27.21 \\
       & Set 2 & \multicolumn{1}{c|}{6.78}  & \multicolumn{1}{c|}{16.61} & 27.85 \\ \Xhline{2\arrayrulewidth}
\end{tabular}
\label{tbl:table2}
\end{table}
We then validate the above claim of the two sets by showing the sensitivity of the performance by inferring input speech with each partition set presented in Table \ref{tbl:table2}. We observe that the WER is statistically unchanged when employing prompt set 1, whereas omitting it would result in increasing error, plausibly caused by the lack of content refinement. Likewise, Fig. \ref{fig:role3} shows that employing prompt set 2 exhibits better noise-variant features compared to prompt set 1, strongly suggesting the role in noise-information-tuning.

\begin{table}[!ht]
\centering\footnotesize
\tabcolsep=0.25cm
\renewcommand{\arraystretch}{1.35}
\vspace{-0.3cm}
\caption{WER of different models, exhibiting the generalization power of the downstream decoder and predictor head by removing the use of all soft prompts during feedforward.}
\begin{tabular}{l|c|ccc}
\Xhline{2\arrayrulewidth}
\multirow{2}{*}{Methods} & \multirow{2}{*}{\begin{tabular}[c]{@{}c@{}}No. of\\ Prompts\end{tabular}} & \multicolumn{3}{c}{WER (\%) of LibriSpeech Test} \\ \cline{3-5} 
                       &                     & \multicolumn{1}{c|}{Clean} & \multicolumn{1}{c|}{Other} & Noisy \\ \hline
HuBERT \ding{100} (Base)               & N.A.                & \multicolumn{1}{c|}{7.09}  & \multicolumn{1}{c|}{17.61} & 29.04 \\ \hline
Prompt Tuning          & \multirow{2}{*}{20} & \multicolumn{1}{c|}{6.64}  & \multicolumn{1}{c|}{16.30} & 27.06 \\
\ding{229} Remove All Prompts &                     & \multicolumn{1}{c|}{6.78}  & \multicolumn{1}{c|}{16.63} & 28.25 \\ \hline
Prompt Tuning          & \multirow{2}{*}{50} & \multicolumn{1}{c|}{6.37}  & \multicolumn{1}{c|}{15.99} & 26.13 \\
\ding{229} Remove All Prompts &                     & \multicolumn{1}{c|}{6.60}  & \multicolumn{1}{c|}{16.39} & 28.37 \\ \Xhline{2\arrayrulewidth}
\end{tabular}
\label{tbl:table3}
\end{table}

Finally, we ask: \textbf{How do the soft-prompts contribute to the optimization of the downstream decoder and predictor head?} Table \ref{tbl:table3} presents the performance of the prompt tuning model when the learned soft prompts are not prepended during utterance inference, resulting in an architecture identical to frozen HuBERT. This table illustrates the optimization process and the generalization capabilities of the downstream decoder and predictor head under the influence of the prompts. While we anticipated a slight increase in recognition errors, it is surprising to observe that performance actually improves compared to the base frozen HuBERT, even without utilizing soft prompts during inference. Notably, the model with a prompt size of 50 exhibits an approximate 7\% gain, indicating improved optimization where the soft prompts guide the latent representations to assist the downstream decoder and predictor head in converging to a better local minimum. As such, we can exploit this property to train a prompt tuning model and remove the soft prompts for zero inferencing cost improvement.

\subsection{Zero-shot Learning on Out-of-Domain Noise}
In this section, we ask: \textbf{Can we use the characterized role of the prompts to achieve zero-shot learning on out-of-domain noise speech?} Previously, we observed that noise prompts improve noisy ASR by imbuing noise information for robust adaptation. We propose to perform domain shift on noise set for zero-transfer on OOD noise environments. This process introduces new inductive biases to facilitate domain adaptation for the current model. In particular, given a sample set of OOD noise audio, we extract acoustic representations from the existing prompt tuning model, pool them globally, and average to obtain a single vector that represents the inductive bias of the OOD noise. We broadcast this vector to match the number of noise prompts, $\mathbf{P_N}$, and apply a Hadamard product for the domain shift to obtain the new noise prompts, $\mathbf{P_N^*}$ (Set 2). The new prompts become $\mathbf{P^* := [P_C, P_N^*]}$, where $\mathbf{P_C}$ refers to the content tuning prompts (Set 1).

\begin{table}[!ht]
\centering\footnotesize
\tabcolsep=0.08cm
\renewcommand{\arraystretch}{1.35}
\vspace{-0.3cm}
\caption{Performance comparison of different models on out-of-domain noise augmented testing sets from LibriSpeech, where noisy speech is simulated with SNR of 0-20dB.}
\begin{tabular}{l|cccc}
\Xhline{2\arrayrulewidth}
\multirow{2}{*}{Methods} & \multicolumn{4}{c}{WER (\%) of Noisy LibriSpeech (0 - 20dB)}         \\ \cline{2-5} 
                         & Dev-Clean & \multicolumn{1}{c|}{Dev-Other} & Test-Clean & Test-Other \\ \hline
HuBERT \ding{100} (Base)                 & 20.78     & \multicolumn{1}{c|}{36.03}     & 19.86      & 35.89      \\ \hline
Prompt Tuning (20)           & 19.67     & \multicolumn{1}{c|}{34.77}     & 18.69      & 34.62      \\
\ding{229} Noise Shifted Prompts & \textbf{19.18} & \multicolumn{1}{c|}{\textbf{33.67}} & \textbf{17.83} & \textbf{33.64} \\ \Xhline{2\arrayrulewidth}
\end{tabular}
\label{tbl:table4}
\end{table}

Table \ref{tbl:table4} uses a separate OOD noise from FSD50K \cite{fonseca2021fsd50k}, where we select office-related background noise to corrupt the official LibriSpeech Dev and Test sets to simulate noisy (0-20dB) audio. We observe around 2.5 to 4.6\% improvement on the evaluating sets, compared to the vanilla prompt tuning model, without utilizing additional computational resources to retrain the model to the new environmental domain. This possesses valuable real-world applicability, which could potentially save costs on ASR adaptation.

\section{Conclusion}
\label{sec:conclusion}
In conclusion, our study has aided the understanding of prompt tuning for ASR. We have found that soft prompts enhance ASR performance, although they are susceptible to malicious modifications, they are not obligatory for inference. Our analysis identified two key roles for soft prompts: content refinement and noise information enhancement, which enhance robustness against background noise. Lastly, we have proposed a modification for noise prompts to allow zero-shot learning adaptation. We hope that our work adds valuable insights to help better appreciate the functions of soft prompts and develop more effective ways to use them.


\vfill\pagebreak

\bibliographystyle{IEEEbib}
\bibliography{refs}

\end{document}